\documentclass[11pt]{article}
\usepackage{moriond,epsfig}

\def\Journal#1#2#3#4{{#1} {\bf #2}, #3 (#4)}

\def\PLB{{\em Phys. Lett.}  B}
\def\PRL{\em Phys. Rev. Lett.}

\def\TKP{\em The KamLAND Prop.}

\def\be{\begin{equation}}
\def\ee{\end{equation}}
\def\bea{\begin{eqnarray}}
\def\eea{\end{eqnarray}}

\usepackage{graphicx}
\begin{document}
\vspace*{4cm}

  \title{An Update on Progress at KamLAND}
  \author{S.A. Dazeley for the KamLAND Collaboration}

  \address{Dept. of Physics and Astronomy \\ 
	Louisiana State University, Louisiana USA 70803\\[2mm]}

\maketitle\abstracts{
The first generation of 
solar neutrino experiments narrowed the allowed 
flavor mixing 
and mass parameter solutions \mbox{(for $\nu_{e} \leftrightarrow \nu_{x}$)} 
to a few isolated
regions of $sin^{2} 2 \theta - \Delta M^{2}$ parameter space.
Recently, the Small Mixing Angle (SMA) solution 
($sin^{2} 2 \theta \sim 10^{-3} \rightarrow 10^{-2}$ and 
$\Delta M^{2} \sim 10^{-5}$ eV$^{2}$), and the
``just so'' ($\Delta M^{2} < 10^{-9}$eV$^{2}$) solutions have been disfavored by
results from Super-Kamiokande~\cite{superk:1}~\cite{smy:2} 
and SNO~\cite{sno:1}.
The Kamioka Liquid scintillator Anti-Neutrino Detector (KamLAND)
recently became operational, and is particularly sensitive to the 
Large Mixing Angle (LMA) region ($sin^{2} 2 \theta \sim 1$ and 
$\Delta M^{2} \sim 10^{-5} \rightarrow 10^{-3}$ eV$^{2}$).
We believe the background impurity levels in the detector are low enough to
conduct a successful experiment.  The stability of
the central balloon and PMTs has also been confirmed.}

\section{Introduction}

The solar neutrino flux measurements of Homestake~\cite{cl:1},
GALLEX~\cite{gallex:1}, SAGE~\cite{sage:1}, Kamiokande~\cite{kamiokande:1}, 
Super-Kamiokande~\cite{superk:1} and SNO~\cite{sno:1}
were
significantly lower than those predicted by Standard Solar Models (SSMs).  
This discrepancy was known as ``the solar neutrino problem'', and
can be explained as a natural consequence of 
neutrino flavor oscillations.
Super-Kamiokande has observed a significantly
lower ratio of $\nu_{\mu}$ to $\nu_{e}$ like events than expected
from neutrinos produced by cosmic ray interactions in the
upper atmosphere.  This was also interpreted as
evidence for neutrino flavor oscillations.
In addition, LSND~\cite{lsnd:1} found evidence for 
$\nu_{\mu} \rightarrow \nu_{e}$ oscillations 
at the Los Alamos Meson Physics Facility.

Solar neutrino experiments are, by definition, long baseline experiments.
They also tend to be low energy.
Most laboratory experiments are short baseline.  This has led to
a lack of experimental coverage of the 
$sin^{2} 2 \theta - \Delta M^{2}$ plane 
near the LMA region.

\section{The KamLAND Experiment}

KamLAND is a medium energy, medium 
baseline reactor anti-neutrino experiment.  It utilizes
the large number of nuclear reactors that lie within a few hundred kilometers
of the KamLAND site, which is situated inside the dome of the old Kamiokande
experiment in Japan.  Assuming no flavor oscillations, 
these reactors produce an average flux (at KamLAND)
of $1 \times 10^{6}$cm$^{-2}$ s$^{-1}$ anti-neutrinos 
at an energy greater than 1.8 MeV \cite{kamland_prop:1}
(the neutrino capture threshold).  In this scenario, 
KamLAND would detect $\sim 2$ per day.  

\begin{figure}[t]
\vspace{7.0cm}
\includegraphics{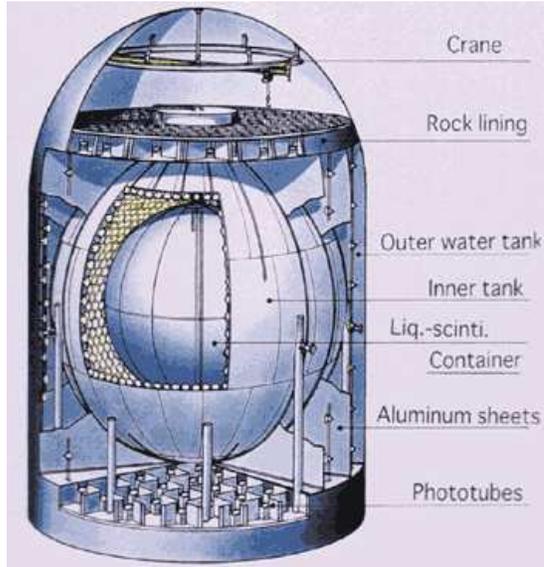}
\caption{A schematic representation of the KamLAND detector.  Shown here
is a cut out of the central balloon, surrounded by a paraffin oil buffer
region contained within a stainless steel sphere
and a water filled outer detector.  The $\sim 2000$ PMTs are shown
on the inside of the stainless steel sphere.}
\label{fig:kamland}
\end{figure}

A schematic of KamLAND is shown in Figure~\ref{fig:kamland}.  The
detector comes in three sections.  The central balloon contains $\sim 1000$
tons of dodecane (80\%), pseudocumene (20\%) and 
diphenyloxazole (PPO) scintillator (1.5 g/L).  
A transparent nylon balloon contains this liquid scintillator 
and is stabilized by a mesh 
of Kevlar ropes suspended from the top (chimney) and bottom of the detector.
Outside the balloon, and also acting as a support, is a mixture of
paraffin oils.  The paraffin oil is contained within an 18 meter diameter
stainless steel
spherical vessel and acts as a buffer region, shielding the balloon volume 
from gamma rays emitted by the Photo-Multiplier Tubes (PMTs) situated
on the inner wall of the vessel.  1878
inward looking 20'' and 17'' PMTs are placed to detect
scintillation light from neutrino interactions inside the balloon.  
The volume inside the stainless steel sphere is called the Inner Detector (ID).
There are 1325
17'' PMTs (which have the same shape as the 20'' PMTs but 
better time resolution), and 553 20'' PMTs  
evenly distributed around the ID.
The ID sits in a cylindrical Outer Detector (OD), which is filled with water.
The OD is used as a veto against
muon events and radioactivity from the rock walls.  On the inner wall 
of the OD are 240 20'' PMTs.

KamLAND can be viewed as a two stage experiment.  The first stage, and
initially the 
most important, is the reactor experiment.  In this phase,
reactor neutrinos are detected via the inverse neutron beta decay reaction
($\bar{\nu}_{e} + p \rightarrow e^{+} + n$).   
The $e^{+}$ is directly observable.  The neutron is captured
on a time scale of $\sim 200$ $\mu$seconds
by a nucleus in the liquid nearby according to the reaction
$n + p \rightarrow d + \gamma (2.2$ MeV).  Therefore the neutrino signature
is a double coincidence in time (delayed) and position.  Additionally, the energy
of the neutron capture must be consistent with a 2.2 MeV gamma.  
The statistical significance of any signal depends upon reducing
the background as much as possible.
To be
successful, low levels of impurity are required in the 
Liquid Scintillator (LS).  The requirements for
important impurities such as U, Th and $^{40}$K are $\sim 10^{-14}$ g/g 
or less, together with low levels of radon.
Preliminary measurements of 
these backgrounds have been performed.  The upper limits for U, Th 
and $^{40}$K are 
$6.4 \times 10^{-16}$ g/g, $1.8 \times 10^{-16}$ g/g,
and $2.3 \times 10^{-16}$ g/g respectively.
Radon levels are much more variable, as they depend on the 
purification system (which is still being improved) and mine air.  
Radon contamination leads to the decay products $^{208}$Tl, 
$^{210}$Pb and $^{210}$Bi, which tend to settle on the balloon surface
and can be monitored by low energy event rate measurements.
We can remove most of this effect by employing a fiducial
volume cut in software on event vertices near the balloon.

The second phase is the
solar neutrino experiment.  KamLAND has the capability to observe 
low energy neutrinos down to
0.6 MeV, enabling the detection of $^{7}$Be solar neutrinos.  
This would require very low impurity levels ($\sim 2$ orders
of magnitude lower than for the reactor phase, i.e. $\sim 10^{-16}$ g/g
or less
of U/Th/$^{40}$K).  While more measurements
need to be done, KamLAND may already be close to 
this goal.

Other possible experiments include using KamLAND 
as a proton decay detector.  The increased light output 
of scintillator over Cherenkov experiments make KamLAND a more efficient
detector of the K$^{+}$ 
than water Cherenkov detectors.  
Simulations suggest
that KamLAND may be almost as effective (for detecting the K$^{+}$ modes)
as Super-Kamiokande, despite its
smaller volume.
Additionally, KamLAND has the capacity to study
atmospheric neutrinos, as well as those produced by
radioactive decay in the Earth's crust and astrophysical sources.

\begin{figure}[t]
\vspace{11.0cm}
\includegraphics{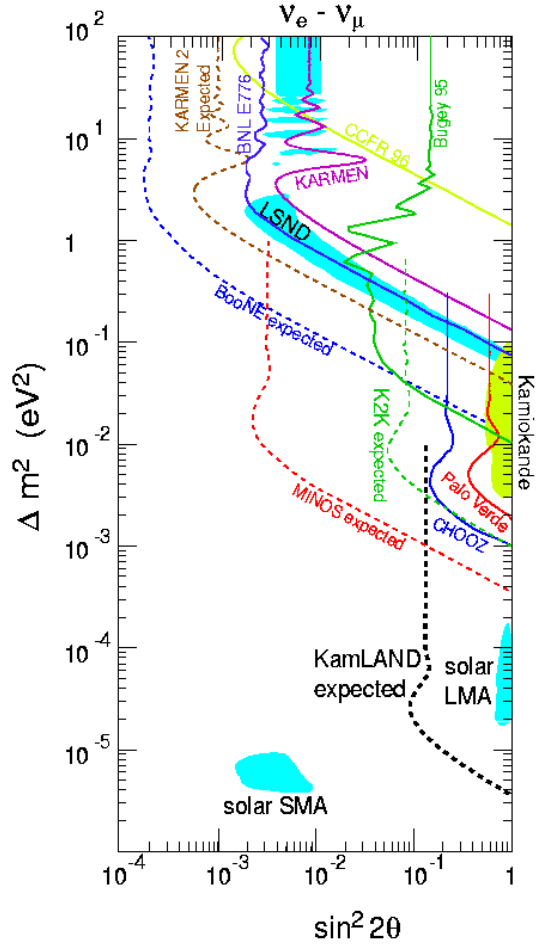}
\caption{A selection of modern neutrino experiments and the regions of 
$\Delta m^{2}$ vs $sin^{2} 2 \theta$ parameter space to which
they are most sensitive.
The figure shows that KamLAND is uniquely sensitive to the LMA region.  
Super-Kamiokande
and SNO have recently published results that disfavor the SMA region.}
\label{fig:kamosci}
\end{figure}

Figure~\ref{fig:kamosci}  shows the regions of 
$\Delta m^{2}$ vs $sin^{2} 2 \theta$ parameter
space KamLAND will be most sensitive to in the reactor phase.  It also shows
where the SMA and LMA solutions are, relative to other experiments.  This 
illustrates why KamLAND is such a unique project.  As it
is the only medium baseline real-time reactor experiment, 
it will be most sensitive to
the LMA solution around $\Delta M^{2} \sim 10^{-4}$ eV$^{2}$ and 
$sin^{2} 2\theta \sim 1$.

\section {Results}

Various forms of calibration are taking place.  Energy calibrations 
will be performed by a number of radioactive sources.  To date, a 
$^{65}$Zn source
at 1.1 MeV energy and a $^{60}$Co source at 2.5 MeV have been used.
Using only the 17'' PMTs,
early results give an average of 240 photo-electrons per MeV. This is 
a little higher than anticipated.  
During the reactor phase, only the fast timing 17'' PMTs are required.
The 553 20'' PMTs have not yet been commissioned.
To check the neutron detection efficiency, a number of Americium/Beryllium
sources which emit a prompt gamma at 4.4 MeV and a fast neutron will be
used.  
For PMT gain calibrations a number of blue 
LEDs have been placed around the periphery
of the detector.  A 337 nm nitrogen laser is also used.  
The light from the laser 
can be sent to the detector through a fiber.  Once it reaches the LS, it
scintillates, simulating a real event.
For time calibrations, a nitrogen dye laser at 530 nm (which doesn't 
scintillate) is used.

Official data taking began at KamLAND on January 22, 2002.
The balloon and PMTs are physically 
stable and the balloon shows no signs of leaking.  
The event rate has averaged 
$\sim 25$ Hz
at a threshold of $\sim 0.8$ MeV.  
The data rate has been $\sim 100$ G-Bytes per day, even at this low event rate. 
This is because flash ADC 
waveforms are recorded for every PMT triggered.

\section{Summary}

KamLAND has been operational since January 22 2002.  
The critical physical elements of the experiment, such as
impurity levels, strength and integrity of the 
central balloon and PMTs have been confirmed.  
We expect that KamLAND will be able to successfully complete
the reactor phase of the experiment, and make 
unique measurements of the LMA region of $\Delta M^{2} - sin^{2} 2 \theta$ 
parameter space.  We hope to be able to show results by late 2002 or
early 2003.

\section*{Acknowledgments}

KamLAND is supported by the Japanese Ministry of Education, 
Culture, Sports, Science and Technology, 
the Japanese Society for the promotion of Science, and 
by the US Department of Energy.


\end{document}